# The Origin of the Exoplanets

HELMUT A. ABT

Kitt Peak National Observatory, Tucson, AZ 85726-6732; abt@noao.edu

**ABSTRACT.** We explore two ways in which objects of planetary masses can form. One is in disk systems like the solar system. The other is in dense clusters where stars and brown dwarfs form. We do not yet have the instrumental accuracy to detect multiplanet systems with masses like those in solar system; with our present technology from a distant site, only the effects of Jupiter could be detected. We show that the orbital characteristics (eccentricities and semimajor axes) of stellar, brown dwarf, and exoplanet companions of solar-type stars are all the same within our measuring accuracies and are very different than the planets in the solar system. The period ratios in multiplanet systems do not distinguish between the two models. We conclude that most of the exoplanets found to date are formed like stellar companions and not in disk systems like the solar system. This conclusion explains why metal-poor stars lack planets: because metal-poor stars lack stellar companions with short periods. The distribution of exoplanetary periods for primaries having [Fe/H] $< -0.3$ fits the distribution for stellar companions of metal-poor stars and not of metal-rich stars.

## 1. INTRODUCTION

There are two ways in which planetary-mass objects can be formed. One way is in stellar disks, in which planets are formed by conglomeration of disk material, as in the solar system. Another is the same way that stars are formed: too much angular momentum is in contracting gas clouds to form single objects, so multiple objects are formed. Nelson et al. (1986) and Boss (2003) have calculated that objects with masses down to planetary masses (1–100 $M_{Jup}$) can be formed in that way. In dense clusters, many double and multiple stars are formed, and many are destroyed, by three-body interactions (Aarseth & Hills 1972).

Initially it was thought that the discovered exoplanets could not be formed like stars because the mass-luminosity relation terminated with the late M-type dwarf stars ($M > 75 \; M_{Jup}$) and there were no brown dwarfs ($13 \; M_{Jup} < M < 75 \; M_{Jup}$). That absence was called the "brown dwarf desert." However, we now know of hundreds of brown dwarfs. The free-floating brown dwarfs are found by direct IR imaging of objects in front of dark clouds. Additionally, brown dwarfs have been found in young clusters such as the alpha Persei cluster (Stauffer et al. 2003) and Pleiades (Zapatero Osorio et al. 1998). Finally, there are many brown dwarfs in the list of exoplanets, namely those with derived minimum masses, $M \sin i$, of 13–25 $M_{Jup}$. Because the main source of radiation in brown dwarfs is deuterium burning that lasts only ∼$2 \times 10^8$ yr, old brown dwarfs will be invisible and can be discovered only as gravitational companions of stars. Probably the number of invisible brown dwarfs can be predicted from the numbers of more massive stars in young clusters.

We do not yet have the radial-velocity sensitivity to detect systems like the solar system. The radial motion of the Sun due to the Earth is only $0.09 \; \mathrm{m \, s^{-1}}$, far less than the current accuracy of 5–10 $\mathrm{m \, s^{-1}}$. The motion due to Jupiter is $12 \; \mathrm{m \, s^{-1}}$ at a distance of 5 AU from the Sun. Observers have discovered at least seven Jupiter-mass objects (55 Cnc d, HD 187123c, HD 217107c, HD 160691e, HD 134987c, 47 UMa d, HD 13931b) at 5 AU from their primaries, but no other exoplanets like the remaining planets in the solar system have been discovered (Schneider 2010). Therefore the bulk of the exoplanets found to date do not have the masses and semimajor axes of those in the solar system. We do not know whether the solar system is typical of disk systems; maybe the solar system harbors habitable planets simply because it is atypical of disk systems.

An uncertainty in the masses of exoplanets is that the radial velocities yield values of $M_2 \sin i$, where $i$ is the angle between the lines of sight and the orbital axes. That angle is known only for transiting systems ($i \sim 90°$), of which 80 are currently known. For the remaining systems, we depend on the evidence that rotational, and probably orbital, axes are distributed randomly (Abt 2001). In random orientation of axes, the mean value of $\sin i$ is $\pi/4 = 0.785$, which is not far from 1 (Chandrasekhar & Munch 1950). Therefore there is little doubt that most of the exoplanets found to date and with $M_2 \sin i < 13 \; M_{Jup}$ have planetary masses. In the current literature there are also 25 systems with known orbital elements and with $13 \; M_{Jup} < M_2 \sin i < 25 \; M_{Jup}$, which are mostly brown dwarfs.

It was shown as early as 1998 by Mayor et al. (1998), who had a sample of 20 exoplanets, or 1/20 of the current sample,

that the orbital elements of the exoplanets are very different than those of the solar system. The exoplanets have large eccentricities and small semimajor axes, unlike the planets in the solar system. We can now strengthen those preliminary results.

## 2. ORBITAL ECCENTRICITIES

Consider first the eccentricities. For a sample of stars, observed mainly with accuracies of $100-1000 \text{ m s}^{-1}$, we consider the F7-K9 IV or V binaries in Batten et al. (1989), from here on referred to as SB8. We excluded the faint ($V > 8$ mag) stars that were first discovered photometrically as eclipsing binaries because we wish to compare this sample with others (brown dwarfs, exoplanets) that were discovered by their radial velocity variations, not by light variations. The distribution of eccentricities of the 188 stars is shown in the top panel of Figure 1. The distribution of eccentricities of brown dwarf companions of solar-type stars and shown in the second panel of Figure 1 come from (a) 15 brown dwarfs in the current listing of exoplanets with $13 \, M_{\text{Jup}} < M_2 \sin i < 75 \, M_{\text{Jup}}$, (b) six brown dwarfs listed in Table 3 of Nidever et al. (2002), and (c) four brown dwarfs in their Table 2 that have orbital elements given in Pourbaix (2004). For the 424 exoplanets discovered by 2010 May 24, we use the 379 with known eccentricities; the distribution is shown in the third panel of Figure 1.

The distributions of eccentricities of the stellar and exoplanet companions are very similar in shape within their estimated errors, which were computed as the square roots of their numbers. We find that the fractions with $e \geq 0.3$ are $28.1\% \pm 3.9\%$ of the stellar companions and $29.0\% \pm 2.7\%$ of the exoplanets. The data for the 25 brown dwarfs are too few to provide a good comparison, but even among those, half have $e \geq 0.3$. Compare that for the solar system, shown in the bottom panel of Figure 1, for which none have $e > 0.2$.

We conclude that the distribution of eccentricities of the exoplanet companions are very much like those of stellar companions and very different than for the planets in the solar system.

## 3. ORBITAL SEMIMAJOR AXES

Additional convincing evidence occurs from the semimajor axes. For a stellar sample we will use the nearby solar-type samples observed by Duquennoy & Mayor (1991) and Abt & Willmarth (2006) because those authors tried to find all the binaries in their samples and therefore show a more realistic distribution of semimajor axes than the SB8 sample. They discovered a total of 64 different binaries, of which 10 have a $\sin i > 4.0$ AU and are off scale in Figure 2 (top panel). For the brown dwarf companions to solar-type stars we used the samples from the current list of exoplanets and from Nidever et al. (2002) described in § 2. Of the 25 systems, three have a $\sin i > 4.0$ AU and are off scale in the figure. For the exoplanets we show the distribution of semimajor axes in the third panel of Figure 2 for 370 objects; an additional 10 have a $\sin i >$

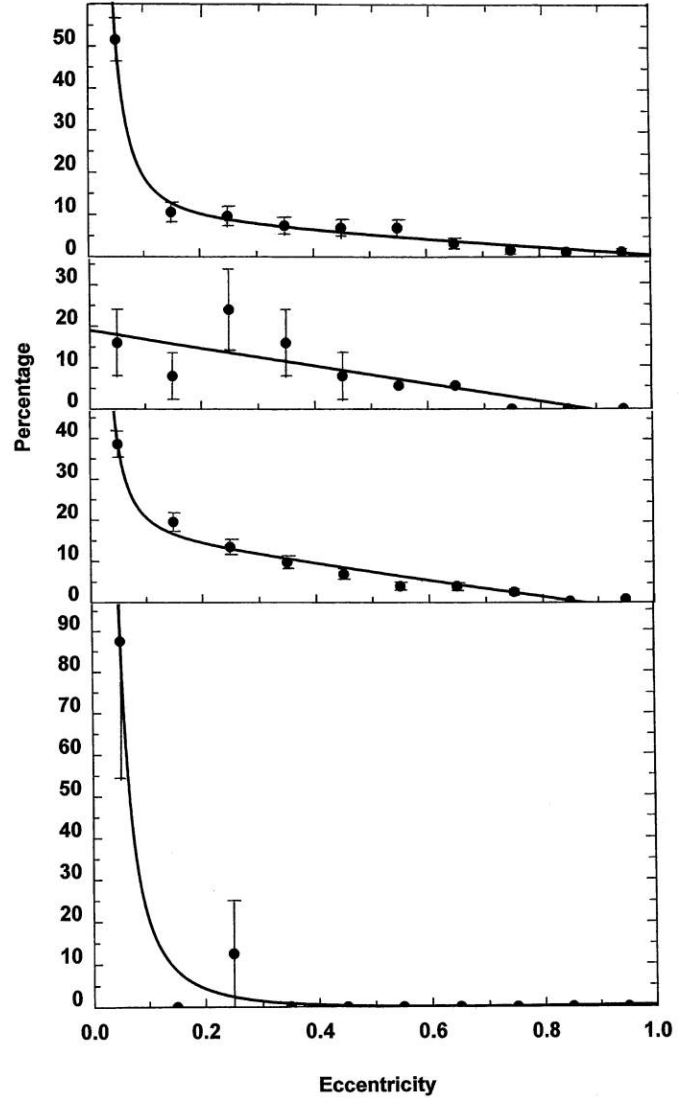

Fig. 1.—Distributions of eccentricities for four types of companions to Sunlike primaries. The *top panel* shows the distribution of eccentricities of the stellar companions of 188 F7-K9 IV or V primaries ($M_2 \sin i > 75 \, M_{\text{Jup}}$) as listed in SB8. The *second panel* shows that for 25 brown dwarf companions with known orbital elements ($13 \, M_{\text{Jup}} \leq M_2 \sin i \leq 75 \, M_{\text{Jup}}$). The *third panel* shows the distributions of eccentricities of exoplanet companions of 379 solar-type primaries ($M_2 \sin i < 13 \, M_{\text{Jup}}$). The *bottom panel* shows the distribution of nonzero eccentricities for the eight planets in the solar system ($M_2 \sin i \leq M_{\text{Jup}}$); the remaining zeros show that the solar system has no eccentricities greater than 0.2″.

4.0 AU and are off scale. The three distributions are the same within their errors.

The distribution of semimajor axes in the solar system is shown in the bottom panel of Figure 2. It shows the four inner planets and the absence of planets to 4.0 AU is shown as zeros. The outer planets are off scale at >4.0 AU. This distribution shows no peak at a $\sin i < 0.05$ AU and nothing between 1.6–4.0 AU, whereas $16.7\% \pm 5.6\%$ of the stellar companions pact (smaller values of a $\sin i$) than the solar system, their planets are unlikely to have the large eccentricities shown by the discovered exoplanets because then those systems would be unstable.

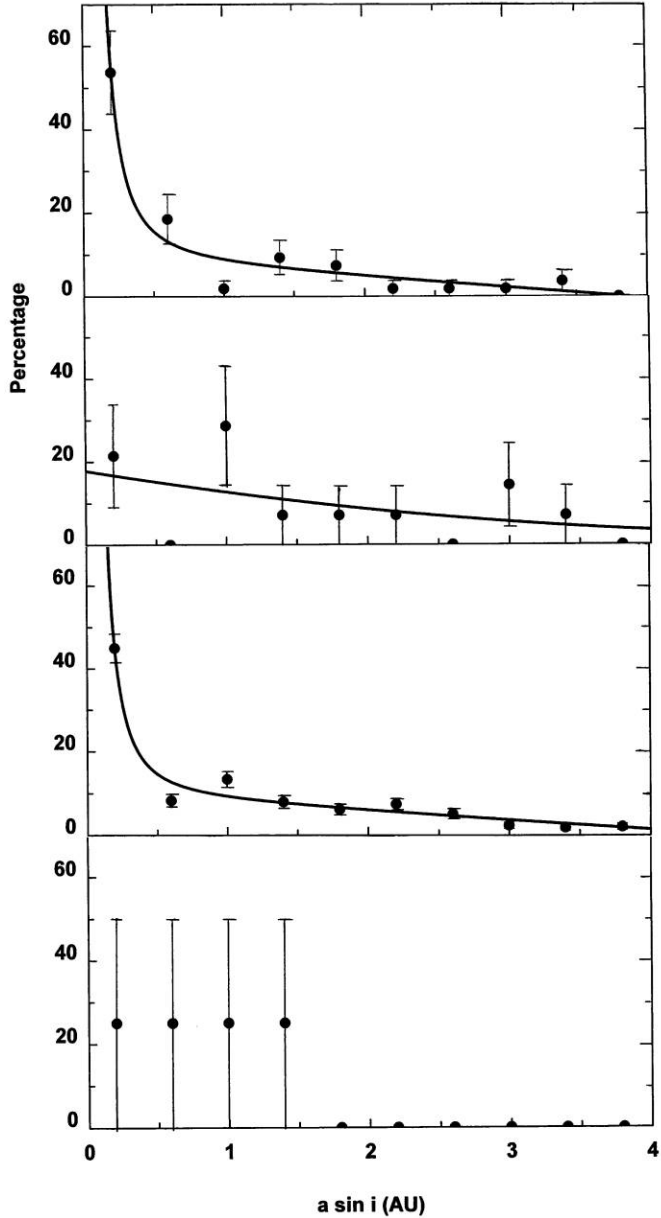

Fig. 2.—Distributions of semimajor axes for four types of companions to Sun-like primaries. The *top panel* shows the distribution of the projected semimajor axes of 54 stellar companions of solar-type primaries ($M_2 \sin i > 75\ M_{\mathrm{Jup}}$). The *second panel* shows that for 21 brown dwarf companions ($13\ M_{\mathrm{Jup}} \lesssim M_2 \sin i \lesssim 75\ M_{\mathrm{Jup}}$). The *third panel* shows that for 327 exoplanets ($M_2 \sin i < 13\ M_{\mathrm{Jup}}$). The *bottom panel* shows that for the four inner planets in the solar system ($M_2 \sin i \leq\ M_{\mathrm{Jup}}$).

## 4. MULTIPLANET SYSTEMS

Triple-star systems invariably have a close pair and a distant third star. Batten (1973) has shown that that is true for both visual and spectroscopic triples. The ratio of the long period to the short one is generally a factor of 10 or more (Tokovinin 2008). On the other hand, in the solar system, the periods of the planets, and of the satellites within planet systems, have progressions with factors from slightly more than 1.0 to 15. The only limitation to their period ratios is the occurrence of resonances.

In the current list of exoplanets, there are 33 systems having two to five exoplanet companions. For the first two companions spatially, the distribution of period ratios of the second and 25.1% ± 2.6% of the exoplanets have semimajor axes in the range of 1.6–4.0 AU. We conclude that the distributions of semimajor axes of the stellar, brown dwarf, and exoplanet companions are the same but differ drastically from that in the solar system, even though the solar system has a typical age for the stars having exoplanets. If typical disk systems are more com-companion ($P_2$) to the first companion ($P_1$) shows a linear decline from factors of 1.1 to 330. That seems to imply that multiplanet systems are more like the solar system than like stellar systems. But that is a wrong conclusion.

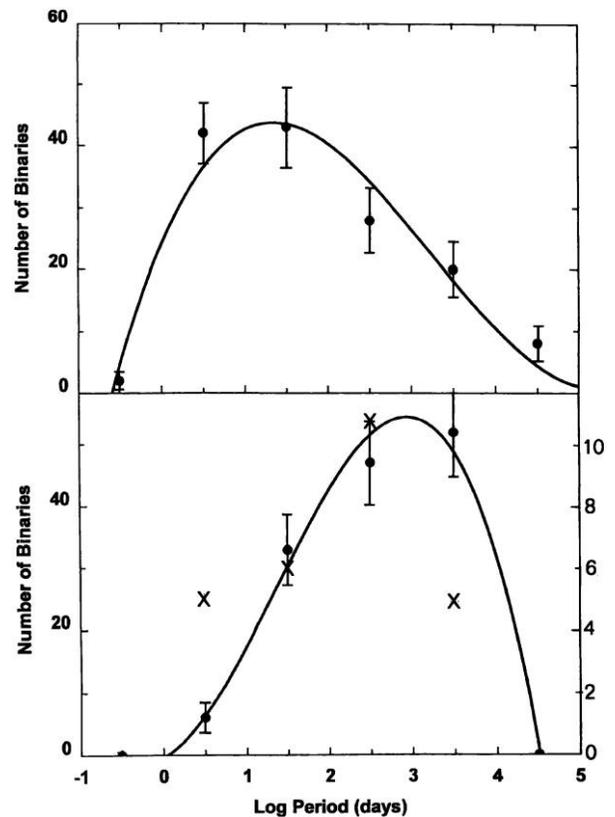

Fig. 3.—*Top panel:* the distribution of periods for 196 FG IV or V stars in Pourbaix's (2004) online compilation and with metal-rich abundances ([Fe/H] > −0.30) is shown. *Bottom panel:* the distribution of periods of 138 late-type stars and with metal-poor abundances ([Fe/H] < −0.30) is shown with *filled circles* and error bars; their vertical scale is on the left. These data are taken from Abt (2008). The data on 27 exoplanets surrounding metal-poor primaries are indicated with x *symbols* and the vertical scale on the right. Their error bars average those of binaries but the bars are not shown to avoid confusion.

TABLE 1
Conclusions Summary

| Criterion | Stellar Formation Model | Disk Model |
|---|---|---|
| Eccentricities .......... | Satisfies | Does not satisfy |
| Semimajor axes ....... | Satisfies | Satisfies only if the solar system is atypical |
| Multiple periods ....... | Satisfies | Satisfies |
| Metal-poor companions ...................... | Satisfies | Does not satisfy |

Most three-body systems are unstable, except in special cases. One stable case involves two objects of similar masses and an object of much smaller mass. That describes the stellar triples. Another stable case involves a single massive object and two or more objects that are several orders of magnitude smaller in mass. That describes the solar system and the exoplanet systems in which the Sun-like primaries have 3 orders of mass greater than the companions. Therefore one should not compare the exoplanet systems that are dominated by a single massive object with stellar systems having several objects of similar mass. A suitable comparison for the exoplanet systems would be the stars orbiting around the central black hole of our Galaxy. Those have a variety of semimajor axes and eccentricities, just as do the exoplanet systems (Lu et al. 2009). Therefore we conclude that the multiexoplanet systems have similar orbital characteristics to stellar systems that are dominated by a single massive object or to disk systems like the solar system. But the similarity to either does not distinguish between the model of exoplanet systems formed in disks those like stellar systems.

## 5. EXOPLANETS AROUND METAL-POOR STARS

The explanation of most exoplanets as low-mass companions formed during star formation, and not disk systems, offers a natural explanation of why few exoplanets have been found around low-metallicity stars (Udry & Santos 2007). The top panel of Figure 3 (Abt 2008) shows the distribution of periods for 196 FG IV or V binaries in Pourbaix's (2004) online compilation, using only those with [Fe/H] > −0.30. I also eliminated the eclipsing binaries that were first discovered by their light variations, rather than by their radial velocity variations. The peak period is 20 days.

The lower panel of Figure 3 shows the period distribution of 138 high proper motion late-type binaries with [Fe/H] < −0.30, discovered by Latham et al. (2002) and Goldberg et al. (2002). The peak period is about 1000 days and very few have short periods, even though short-period binaries have the largest velocity amplitudes and are the easiest to discover.

The data for the 27 known exoplanetary companions to stars with [Fe/H] < −0.3 (x symbols, with scale to the right), with errors of about ±2.5 binaries, fit the lower curve in Figure 3 up to periods of about 1000 days (log period = 3); observing times for exoplanets have not allowed the discovery of many longer periods. If one wishes to find more exoplanets around metal-poor stars, one should look for periods of years and decades.

Therefore, if most of the known exoplanets were formed as stellar companions, there should be few around metal-poor stars, and that is what has been observed.

## 6. CONCLUSIONS

The results from §§ 2–5 are summarized in Table 1. We see that the stellar formation model satisfies all four criteria that we considered but that the disk model fails at least two of them, based on the characteristics of the 424 exoplanets discovered to date. This implies that most of those exoplanets were formed in star formation at the low end of the mass-luminosity relation, not in stellar disks. This scenario was proposed on theoretical grounds by Reipurth & Clarke (2001) and Delgado-Donate & Clarke (2003).

We do not yet have the radial velocity sensitivity to detect disk systems like the solar system, except for single large planets, like Jupiter at 5 AU. One can argue that the solar system is not typical of most disk systems and that we, in a disk system with only one habitable planet, can explore this question because Earth-like planets at 1 AU have not yet been discovered. A final test will come when we have three-dimensional data of multiexoplanet systems. Stellar systems tend to be spherical in their distribution of components (Fekel 1981), while disk systems are flat or two-dimensional. However, highly eccentric orbits do not seem to occur in disk systems, and the evidence that metal-poor primaries do not have short-period exoplanets strongly suggests that the exoplanets discovered to date were mostly formed like low-mass stellar companions, not in disks.